\documentclass[11pt]{article}
\usepackage{epsfig}
\usepackage{latexsym}
\usepackage{named}
\usepackage{hyperref} 
\usepackage{lscape}
\providecommand{\citep}{\cite}
\providecommand{\citet}{\cite}

\catcode`\.=\active\def.{\char'56\allowbreak}\catcode`\.=12
\catcode`\/=\active\def/{\char'57\allowbreak}\catcode`\/=12
\catcode`\:=\active\def:{\char'72\allowbreak}\catcode`\:=12  
\catcode`\@=\active\def@{\char'100\allowbreak}\catcode`\@=12 
\catcode`\-=\active\def-{\char'55\allowbreak}\catcode`\-=12  
\def\URL{\bgroup\catcode`\.=\active\catcode`\/=\active\catcode`\:=\active\catcode`\@=\active\catcode`\-=\active\def~{\char126}\tt\URLaux}
\def\URLaux#1{#1\egroup}

\newlength{\halfwidth}
\newlength{\temp}
\newlength{\tempa}

\begin{document}

\title{\vspace*{-1in}\bf
Automated DNA Motif Discovery}


\author{
\protect\href{http://www.cs.ucl.ac.uk/staff/W.Langdon}
{W. B.~Langdon},
Olivia {Sanchez Graillet},
A. P. Harrison
\\[2ex]
Department of Computer Science,
King's College London
\\[1ex]
Departments of Mathematical Sciences and Biological Sciences  \\
University of Essex, UK
}



\bibliographystyle{named}

\date{} 

\maketitle

\begin{abstract}
\normalsize
\mbox{Ensembl}'s human non-coding and protein coding genes
are used to 
automatically find
DNA pattern motifs.
The Backus-Naur form (BNF) grammar for regular expressions 
is used by genetic programming
to ensure the generated strings are legal.
The evolved motif suggests
the presence of Thymine followed by one or more Adenines etc.\
early in transcripts
indicate a non-protein coding gene.
\end{abstract}

\noindent
Keywords:
pseudogene, short and microRNAs, non-coding transcripts,
systems biology,
machine learning,
strongly typed genetic programming 

\section{Introduction}

We present a new method for finding DNA motifs.
First we will describe the existing work 
which uses grammars to
constrain the artificial evolution of programs and its application
to finding patterns,
particularly
finding protein motifs.
The Methods section~(\ref{sec:Methods})
describes how Ensembl~\cite{Ensembl:2009}
DNA sequences are prepared and used.
The new grammar based genetic programming (\ref{sec:GP})
is demonstrated (Section~\ref{sec:results})
by its ability to automatically find patterns early in
human genes which distinguish non-protein coding genes
genes from protein coding genes.

\subsection{Evolving Grammars and Protein Motifs}


Existing research on using grammars to constrain the artificial evolution of
programs can be broadly divided in two:
Grammatical Evolution \citep{oneill:2001:TEC}
which uses BNF grammars and is
based largely in Ireland
and work in the far east
using context-free grammars, tree adjoining grammars and inductive logic
by Whigham,
McKay
and Wong.
See, for example,
\citet{whigham:1996:sblbGP,whigham:1999:aigp3},
\cite{Hoang:2008:IJKBIES} 
and~%
\cite{wong:1996:aigp2}. 
Grammars are also used in many Bioinformatics applications,
particularly dealing with sequences.

Ross
induced stochastic regular expressions from a
number of grammars to classify proteins from their amino acid
sequence \citep{ross:2001:gecco}.
Regular expressions have been evolved 
to search for similarities
between proteins,
again based on their amino acid sequences~%
\citep{oai:biomedcentral.com:1471-2105-8-23}. 
Whilst Brameier 
used
amino acids sequences to predict the location of proteins
by applying a multi-classifier~\citep{langdon:2001:eROC}
linear genetic programming (GP) based approach
\citep{NucPred-bioinformatics2007}
(although this can be done without a grammar \citep{langdon:2005:CS}).
A similar technique has also been applied to study microRNAs
\citep{Brameier:2007:BMCbinf}.
An interesting departure is Pappa's work
which uses a grammar based GP to create application domain specific
algorithms.
E.g.\ \cite{pappa:2009a},
which considers prediction of protein function.
While Dyrka and Nebel have
used a genetic algorithm and 
a more powerful but also more complicated context free grammar.
For example, they used a CFG when
finding a meta-pattern
describing protein sequences
associated with zinc finger RNA binding sites
\cite{Dyrka:2009:bmcBI}.
Zinc finger was amongst the protein superfamilies sequence prediction 
tasks used by
\cite{Dobson:2009:JIB}.
Although Support Vector Machines can achieve high accuracy
(they obtained 66.3\%)
SVM models can be difficult for non-specialists to understand.

Non-stochastic machine learning techniques have also been applied to DNA motifs.
E.g.\
\citep{hu:2000:BI}, present a method based on decision trees,
specifically C4.5.
Note we are deliberately seeking intelligible motifs and so
rule out approaches, such as
\cite{Won:2007:bmcBI},
which evolved high performance but non-intuitive models
for protein secondary structure prediction.
\cite{George:2009:MBT} 
concisely list current computational techniques 
used with RNA motifs.

We must be wary of over claiming.
As 
\cite{Baird:2006:RNA}
point out computational prediction is hard.
Indeed they say for one problem
(identification of new
internal ribosome entry sites (IRES) in viral RNA)
it is still not possible.
Nevertheless,
by concentrating on a generic tool which generates 
human readable motifs,
of a type which are well known to Biologists,
computers may still be of assistance.

\section{Methods} 
\label{sec:Methods}
\subsection{Preparation of Training Data}
\label{sec:inputs}

The DNA sequences for all human genes were taken from Ensembl
(version~48).
There are 46\,319 protein coding
and 9\,836 non-coding transcripts.
(Many genes have more than one transcript.
There are 22\,740 coding and 9\,821 non-coding human genes.)
As Table~\ref{tab:non_coding} shows most non-protein coding human genes
are either pseudogenes of some sort or lead to
short or micro-RNAs.

\begin{table}
\caption{\label{tab:non_coding}
Number and type of each non-protein coding Ensembl human gene}
\begin{center}
\begin{tabular}{lr}
           pseudogene & 1516\\
                snRNA & 1337\\
            misc\_RNA & 1041\\
                miRNA & 968\\
    scRNA\_pseudogene & 843\\
               snoRNA & 716\\
 Mt\_tRNA\_pseudogene & 603\\
      retrotransposed & 565\\
    snRNA\_pseudogene & 501\\
   snoRNA\_pseudogene & 486\\
     rRNA\_pseudogene & 341\\
                 rRNA & 334\\
           V\_segment & 236\\
     tRNA\_pseudogene & 129\\
           J\_segment & 99\\
           C\_segment & 36\\
           D\_segment & 32\\
             Mt\_tRNA & 22\\
    miRNA\_pseudogene & 21\\
misc\_RNA\_pseudogene & 7\\
             Mt\_rRNA & 2\\
                scRNA & 1\\
\multicolumn{1}{r}{total} & 9836
\end{tabular}
\end{center}
\end{table}

We need to be able to check later 
that the automatically generated motif is general.
I.e.\ it has not
over fitted the examples it has seen
and does not fail on new unseen examples.
Therefore
the protein coding and non-coding genes were randomly split in half.
(Transcripts for the same gene were kept together).
One half is available for training the GP and the second is never seen by
GP and is reserved for demonstrating the performance of the evolved motif.
The training data were then processed for use by the GP.

\subsubsection{Training Data Sets for Generating DNA Motifs}

Where a gene has multiple transcripts one was randomly chosen
to be included in the training data. 
The other transcripts for the same Ensembl gene were not used for
training.

\begin{figure}
\centerline{\includegraphics{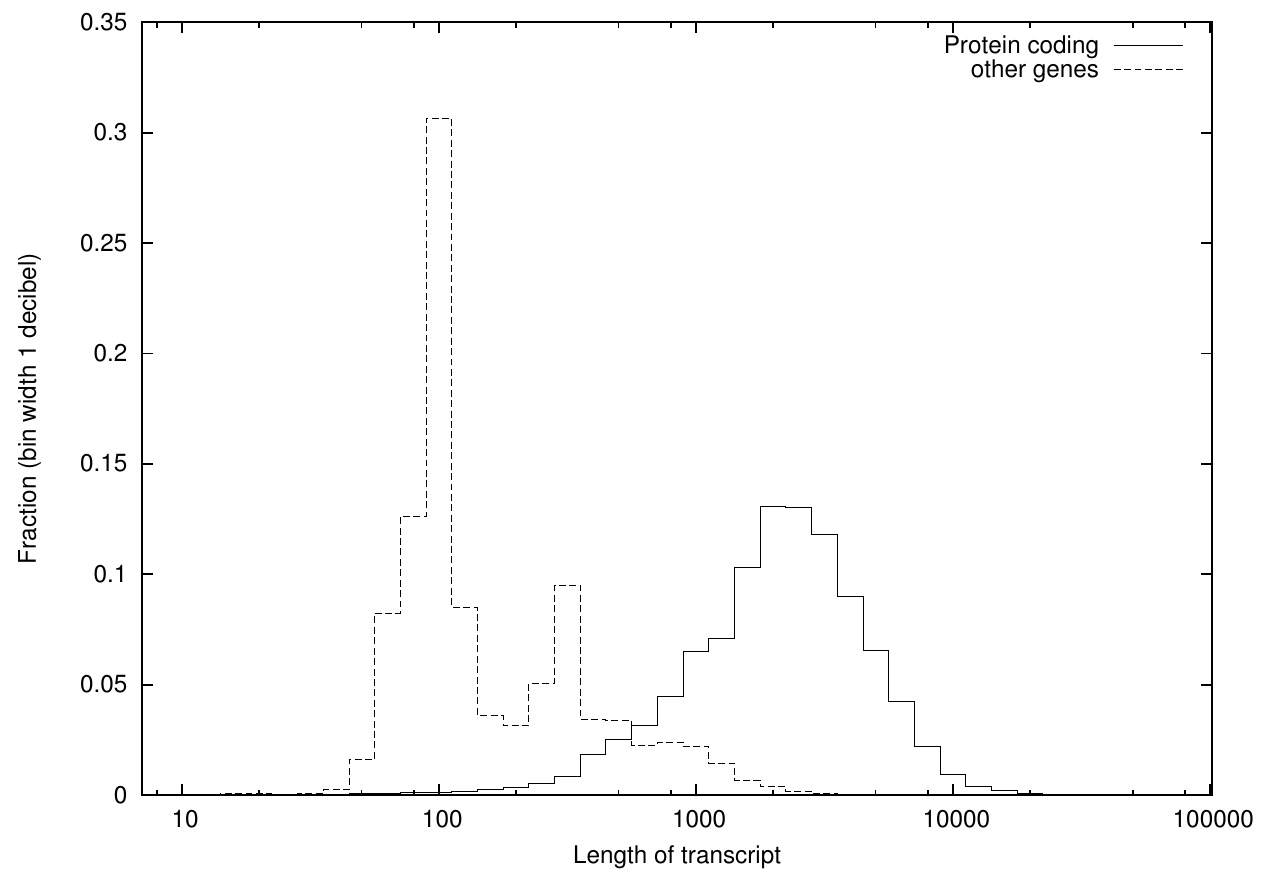}} 
\caption{Distributions of number of bases per Ensembl human transcript.
Note the length of protein coding transcripts is approximately log normally
distributed.
Most non-coding genes are shorter than protein coding genes.
}
\label{fig:gene_length}
\end{figure}

Figure~\ref{fig:gene_length} makes it clear that transcripts from
non-coding genes
tend to be shorter than those produced by protein coding genes.
If the length of the transcript is known,
this would be a very easy way to distinguish protein coding genes.
However a classifier which simply said
``if the transcript exceeds 500 bases, the gene encodes a protein''
would tell us nothing new
(even though it might be quite good at predicting).
So we insist the GP seek out
predictive DNA sequences.
Therefore the GP is not told how long the transcript is.
Instead 
all training data have exactly 60 bases taken from the
start of the Ensembl transcript.
(Transcripts less than 60 bases were 
not used for training).
Finally duplicate sequences were removed.
This gave 4639 unique non-protein coding 
and 11\,191 unique protein coding 
sequences for use as training examples.

\subsubsection{Genetic Programming Training Set}
\label{sec:training_set}

To avoid unbalanced training sets,
every generation all 4639 non-protein coding examples were used
and 4639 coding examples were randomly chosen from the 11\,191
protein coding examples available.
This is done by placing the coding examples at random in a ring.
Each generation the next 4639 examples are taken from the ring.
This ensures the coding examples are regularly re-used.
(Each protein coding example is used once per 2.41 generations.)

\subsection{Evolving DNA Motifs}
\label{sec:GP}

Having created training data we then use a
strongly typed tree GP system
\citep{poli08:fieldguide}
to create an initial random population of motifs.
Each generation the best 20\% are chosen
and a new generation of motifs is created from them using
two types of mutation (shrink and subtree)
and subtree crossover 
\citep{poli08:fieldguide,langdon:book}.
(The exact parameters are given in Table~\ref{gp.details}.)
Over a number of generations the performance of the best
motifs in the population improves.
After 50~generations we stop the GP
and take the best motif at that point
and see how well it does.
It is not only tested on the DNA sequences
used to train it but,
in order to estimate how well it does in general,
it is tested also on the DNA sequences
kept back
(cf.\
Section~\ref{sec:inputs}).

\subsubsection{Backus-Naur Form Grammar of Motifs}
\label{sec:BNF}


The
BNF grammar is given in \cite[Figure~8, page~10]{langdon:2009:AMB}.
Whilst it
could be tuned to each application,
this has not been necessary.
In fact,
we have used the same grammar for a very different task
(isolating poorly performing Affymetrix cDNA probes
\cite{langdon:2009:AMB}).
Technical details and the reasons for its design are given in
\cite{langdon:2009:AMB} 
and
\citet{langdon:2008:CES-483}.


The initial population of motifs is created by 
passing at random through the BNF grammar
using the standard GP algorithm 
(ramped half-and-half \citep{poli08:fieldguide}).
Although this may seem complex,
\verb|gawk| (an interpreted language)
is fast enough to handle populations of a million individuals.

\subsubsection{Creating New Trial Motifs}
\label{sec:xoverBNF}

After each generation,
the best 20\% of the current population are chosen to be the parents
of the next generation.
Each parent is allocated (on average) five children.
Thus the next generation is the same size as the previous one.

Children are created by either mutating high scoring parents
or by recombination of two high scoring parents,
cf.~\citep[Figure~2.5]{poli08:fieldguide}.
In all cases the changes are made so
that the resulting offspring obeys the BNF syntax rules and so
are valid motifs. Therefore their performance can be estimated
and (although some may perform badly)
they are all still comprehensible motifs.

\subsubsection{Evaluating the Motifs}

Each generation 
each trial motif in the population is tested against the
DNA sequences of the 4639 unique non-protein coding 60 base sequences
available for training
and
4639 protein coding
60 base unique sequences selected for use in this generation.
Their performance is the sum of the number of
non-coding sequences they match
and the number of protein coding they do not match.
%
However motifs
which either match all or fail to match any
are penalised by subtracting 4639 from their score.


\section{Results}
\label{sec:results}

\begin{figure}
\centerline{\includegraphics{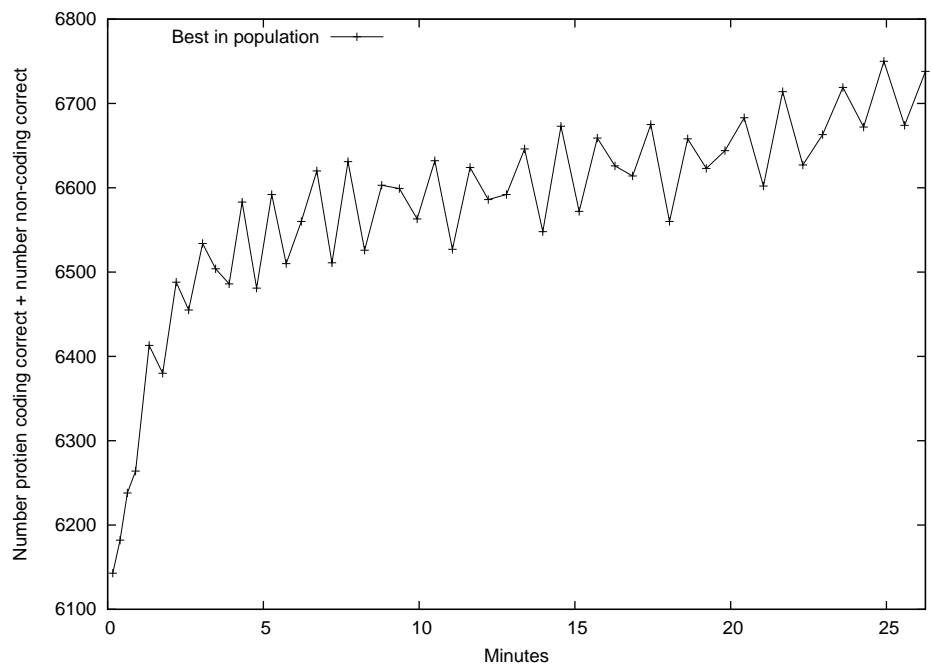}}
\caption{
Evolution of breeding population
of motifs trying to locate human protein coding genes.
Each generation the protein coding training cases are replaced
leading to fluctuations in the measured performance.
However the trend is steadily upwards.
}
\label{fig:p1000_pops}
\end{figure}

At the end of the first run,
with a population of 1000
(cf.\ Table~\ref{gp.details} and
Figure~\ref{fig:p1000_pops})
genetic programming produced the motif
\verb'TACT|TGAT..|TA+TAT.|TA+(.CA+|T)(C|T)'.
(This motif can be understood by noting the vertical bar \verb'|' 
indicates options.
That is, if a sequence contains
\verb'TACT' or \verb'TGAT..' or \verb'TA+TAT.' or \verb'TA+(.CA+|T)(C|T)'
the motif is said to match it.
The last two vertical bars are inside brackets \verb'()'
which must be taken into account before the vertical bar they enclose.
Thus the \verb'(C|T)' means either a Cytosine or a Thymine
placed immediately after bases which match \verb'TA+(.CA+|T)'.
The dots ``\verb'.''' mean one of the four bases must occur here.
Finally \verb'A+' means a run of at least one Adenines.

\begin{table}
\setlength{\temp}{\textwidth}
\settowidth{\tempa}{Performance:}
\addtolength{\temp}{-\tempa}
\addtolength{\temp}{-2\tabcolsep}
%
\caption{\label{gp.details}
Strongly Typed Grammar based GP
Parameters for Pseudogene and
\mbox{non-coding} short RNA Prediction
}
\begin{tabular}{@{}lp{\temp}@{}}\hline
Primitives: \rule[1ex]{0pt}{6pt} & 
The functions and inputs and how they are combined
is defined by the BNF grammar
\protect\cite[Figure~8, page~10]{langdon:2009:AMB}
\\
Performance:       & true positives$+$true negatives.
(I.e. proportional to the area under the ROC curve or Wilcox statistic
\citep{langdon:2004:ECDM}.)
Less large penalty if 
it matches all RNA training sequences or none.
\\ 
Selection:     & (200,1000)
generational, non-elitist, Population size = 1000 
\\ 
Initial pop: & Ramped half-and-half 3:7
\\
Parameters:    & 
90\% subtree crossover,
5\% subtree mutation,
5\% shrink mutation.
Max tree depth~17   
(no tree size limit) 
\\ 
Termination:   & 50 generations
\\ \hline
\end{tabular}
\end{table}

\pagebreak[4]
Confusion matrices are a compact way to show the performance
of prediction algorithms.
They are particularly useful 
where there are many more examples of one class
(e.g.\
protein coding)
than another.
An inept classifier which always said ``protein coding''
would often be correct and so have a high percentage accuracy.
However it would be useless.
By showing how well it does on all types of transcript
a confusion matrix reveals its real performance.
The matrix says how well the classifier does
on each actual class (the columns).
Where there are many classes,
confusion matrices can also be helpful by showing where the
classifier's predictions (the rows) are wrong.
An good classifier will have a matrix with high values 
only on its leading diagonal.

The following pair of confusion matrices give the evolved motif performance on 
its own training data
(i.e.~the training data used in the last generation)
and on all the data used by GP\@.
Of course the actual non-coding examples are the same in the two cases.
However the motif performs equally well on all the protein coding training examples
as it does on the protein coding examples 
randomly selected for us in the last generation.
(I.e.\ they are not significantly different, $\chi^2$, 1~dof.)
This suggests the strategy of randomly changing training examples
every generation has worked well.

\begin{center}
\begin{tabular}{cc}
Last GP generation & GP training data\\
\begin{tabular}{l|rr|rr|}
\multicolumn{1}{c}{} &\multicolumn{2}{c}{non protein   } &\multicolumn{2}{c}{protein coding} \\
\cline{2-5}
non protein    & 3483 & (75\%) & 1403 & (30\%)\\
protein coding & 1156 & (25\%) & 3236 & (70\%)\\
\cline{2-5}
\end{tabular}
&
\begin{tabular}{|rr|rr|}
\multicolumn{2}{c}{non protein   } &\multicolumn{2}{c}{protein coding} \\
\cline{1-4}
 3483 & (75\%) & 3390 & (30\%)\\
 1156 & (25\%) & 7801 & (70\%)\\
\cline{1-4}
\end{tabular}
\end{tabular}
\end{center}

The next pair of confusion matrices
are included for completeness.
The  left hand side
gives
the evolved motif's performance on the 
first 60 bases of the
whole of the training
data
(i.e.\ including duplicates).
The right hand confusion matrix refers to when the evolved pattern is applied
to the whole transcript,
rather than just its first 60 bases.

\begin{center}
\begin{tabular}{cc}
All training data (60) & All training data (whole transcript)
\\
\begin{tabular}{l|rr|rr|}
\multicolumn{1}{c}{} &\multicolumn{2}{c}{non protein   } &\multicolumn{2}{c}{protein coding} \\
\cline{2-5}
non protein    & 3572 & (75\%) & 3447 & (30\%)\\
protein coding & 1196 & (25\%) & 7899 & (70\%)\\
\cline{2-5}
\end{tabular}
&
\begin{tabular}{|rr|rr|}
\multicolumn{2}{c}{non protein   } &\multicolumn{2}{c}{protein coding} \\
\cline{1-4}
 4535 & (92\%) &11227 & (99\%)\\
  375 & (8\%) &  143 & (1\%)\\
\cline{1-4}
\end{tabular}
\end{tabular}
\end{center}

The next pair of confusion matrices contain the evolved
motif's performance on all the holdout data 
(selecting only one transcript per gene).

\begin{center}
\begin{tabular}{cc}
Holdout data (60) & Holdout data (whole transcript)
\\
\begin{tabular}{l|rr|rr|}
\multicolumn{1}{c}{} &\multicolumn{2}{c}{non protein   } &\multicolumn{2}{c}{protein coding} \\
\cline{2-5}
non protein    & 3609 & (76\%) & 3503 & (31\%)\\
protein coding & 1159 & (24\%) & 7844 & (69\%)\\
\cline{2-5}
\end{tabular}
&
\begin{tabular}{|rr|rr|}
\multicolumn{2}{c}{non protein   } &\multicolumn{2}{c}{protein coding} \\
\cline{1-4}
 4529 & (92\%) &11207 & (99\%)\\
  382 & (8\%) &  163 & (1\%)\\
\cline{1-4}
\end{tabular}
\end{tabular}
\end{center}

The last pair of matrices include all transcripts for each of the hold out
genes.
The motif holds its performance when applied to the first 60 bases 
of each Ensembl transcript.
However the shortness of the motif and 
the fact it can match the transcript at any point
means the start of the transcript must be
selected before using the motif
otherwise performance falls.
(Cf.\
the right hand of the previous two pairs of confusion matrices
and the right hand of next pair.)

\begin{center}
\begin{tabular}{@{}cc}
Holdout data (all transcripts, $\le60$) & Holdout (all transcripts, whole transcript)
\\
\begin{tabular}{@{}l|rr|rr|}
\multicolumn{1}{@{}l}{}&\multicolumn{2}{c}{non protein   } &\multicolumn{2}{c}{protein coding} \\
\cline{2-5}
non protein    & 3683 & (75\%) & 6883 & (30\%)\\
protein coding & 1234 & (25\%) &16101 & (70\%)\\
\cline{2-5}
\end{tabular}
&
\begin{tabular}{|rr|rr|}
\multicolumn{2}{c}{non protein   } &\multicolumn{2}{c}{protein coding} \\
\cline{1-4}
 4541 & (92\%) &22778 & (99\%)\\
  376 & (8\%) &  206 & (1\%)\\
\cline{1-4}
\end{tabular}
\end{tabular}
\end{center}

Unlike many machine learning applications,
there is no evidence of over fitting.
Indeed
the corresponding results for the holdout set
are not significantly different \mbox{($\chi^2$, 3~dof)}
from those on the whole training set.
(Both when looking at the first 60 bases or the whole transcript).

Table~\ref{tab:test_p1000} gives a break down of the evolved regular
expression motif both by Ensembl human transcript type and by its
components.
(Note \verb!TA+(.CA+|T)(C|T)!) has been re-expressed as
the union of four expressions:
\verb!TA+.CA+C!, \mbox{ }
\verb!TA+.CA+T!, \mbox{ }
\verb!TA+TC! and
\verb!TA+TT!.)
The last part of the motive
(i.e.\ \verb!TA+(.CA+|T)(C|T)!) 
typically scores more highly than the first three.
However the evolved pattern succeeds 
at separating the non-protein coding from
the protein genes by working together.

It is sufficient for just one of the seven patterns to match the beginning
of the gene.
In many cases either several of the seven match and/or they match the
DNA more than once.
However the patterns are usually  distinct in that,
even in a gene which is matched by more than one of the 7 patterns,
a part of the DNA which matches one is unlikely to also match another.

Although 
the evolved motif has some similarity with the TATA box motif,
it does not match the consensus sequence
\verb'TATAAA'~\cite{Yang200752} exactly.
\verb'TATAAA' occurs in the first 60 bases
in 1.1\% (106) of the 9\,836 non-protein transcripts
and 0.6\% (290) of the 46\,319 protein transcripts.
Depending on the expected prevalence of the four bases,
this is
about what would be expected
by chance.

\section{Discussion}
\label{sec:discuss}


The combination of genetic programming and a BNF grammar
designed for the production of intelligible patterns
can be a viable way to automatically find 
interesting motifs in DNA and RNA sequences.
The prototype system is available via
\href{http://www.cs.ucl.ac.uk/staff/W.Langdon/ftp/gp-code/RE_gp.tar}
{\url{ftp://cs.ucl.ac.uk/genetic/gp-code/RE\_gp.tar}}.)
It has been demonstrated on a large biological DNA problem:
discriminating non-protein coding from protein coding genes.

The automatically generated motif 
\verb'TACT|TGAT..|TA+TAT.|TA+(.CA+|T)(C|T)'
suggests that 
Thymine followed by one or more Adenine bases
(particularly if the run 
is terminated by another Thymine or
a Cytosine and Thymine)
at the start of a transcript,
indicates the transcript
may be a short non-coding RNA sequence
rather than from a protein-coding gene.

\section*{Acknowledgement}

This work was supported by the UK
Biotechnology and Biological Sciences Research Council.
under grant code BBSRC BBE0017421.

\begin{landscape}
\begin{table}
\caption{\label{tab:test_p1000}
Performance of motif on first 60 bases
by components and Ensembl 
transcript type
}
\footnotesize
\begin{tabular}{|l|rr|rr|rr|rr|rr|rr|rr||rr|}
\multicolumn{1}{c}{transcript type} &\multicolumn{2}{c}{\tt TACT} &\multicolumn{2}{c}{\tt TGAT..} &\multicolumn{2}{c}{\tt TA+TAT.} &\multicolumn{2}{c}{\tt TA+.CA+C} &\multicolumn{2}{c}{\tt TA+.CA+T} &\multicolumn{2}{c}{\tt TA+TC} &\multicolumn{2}{c}{\tt TA+TT} &\multicolumn{2}{c}{Combined}\\
\hline
pseudogene &
158 & (10\%) & 269 & (17\%) & 41 & (2\%) & 67 & (4\%) & 42 & (2\%) & 171 & (11\%) & 196 & (12\%) & 676 & (44\%)  \\
snRNA &
739 & (55\%) & 448 & (33\%) & 109 & (8\%) & 120 & (8\%) & 753 & (56\%) & 217 & (16\%) & 737 & (55\%) & 1237 & (92\%)  \\
misc\_RNA &
166 & (15\%) & 671 & (64\%) & 18 & (1\%) & 55 & (5\%) & 73 & (7\%) & 389 & (37\%) & 429 & (41\%) & 992 & (95\%)  \\
miRNA &
197 & (20\%) & 161 & (16\%) & 154 & (15\%) & 42 & (4\%) & 64 & (6\%) & 102 & (10\%) & 327 & (33\%) & 652 & (67\%)  \\
scRNA\_pseudogene &
463 & (54\%) & 157 & (18\%) & 12 & (1\%) & 36 & (4\%) & 52 & (6\%) & 131 & (15\%) & 134 & (15\%) & 671 & (79\%)  \\
snoRNA &
142 & (19\%) & 395 & (55\%) & 75 & (10\%) & 43 & (6\%) & 96 & (13\%) & 144 & (20\%) & 212 & (29\%) & 588 & (82\%)  \\
Mt\_tRNA\_pseudogene &
68 & (11\%) & 179 & (29\%) & 52 & (8\%) & 72 & (11\%) & 125 & (20\%) & 168 & (27\%) & 235 & (38\%) & 518 & (85\%)  \\
retrotransposed &
69 & (12\%) & 75 & (13\%) & 4 & (0\%) & 23 & (4\%) & 26 & (4\%) & 66 & (11\%) & 57 & (10\%) & 237 & (41\%)  \\
snRNA\_pseudogene &
201 & (40\%) & 210 & (41\%) & 34 & (6\%) & 82 & (16\%) & 169 & (33\%) & 66 & (13\%) & 208 & (41\%) & 465 & (92\%)  \\
snoRNA\_pseudogene &
121 & (24\%) & 301 & (61\%) & 23 & (4\%) & 7 & (1\%) & 93 & (19\%) & 26 & (5\%) & 94 & (19\%) & 437 & (89\%)  \\
rRNA\_pseudogene &
39 & (11\%) & 176 & (51\%) & 4 & (1\%) & 98 & (28\%) & 39 & (11\%) & 55 & (16\%) & 35 & (10\%) & 263 & (77\%)  \\
rRNA &
28 & (8\%) & 285 & (85\%) & 2 & (0\%) & 222 & (66\%) & 24 & (7\%) & 52 & (15\%) & 7 & (2\%) & 320 & (95\%)  \\
V\_segment &
35 & (14\%) & 26 & (11\%) & 6 & (2\%) & 7 & (2\%) & 2 & (0\%) & 17 & (7\%) & 41 & (17\%) & 89 & (37\%)  \\
tRNA\_pseudogene &
10 & (7\%) & 32 & (24\%) & 3 & (2\%) & 6 & (4\%) & 11 & (8\%) & 26 & (20\%) & 15 & (11\%) & 77 & (59\%)  \\
J\_segment &
17 & (17\%) & 15 & (15\%) & 1 & (1\%) & 10 & (10\%) & 5 & (5\%) & 18 & (18\%) & 22 & (22\%) & 60 & (60\%)  \\
C\_segment &
0 & (0\%) & 4 & (11\%) & 0 & (0\%) & 4 & (11\%) & 0 & (0\%) & 3 & (8\%) & 3 & (8\%) & 8 & (22\%)  \\
D\_segment &
5 & (15\%) & 2 & (6\%) & 0 & (0\%) & 0 & (0\%) & 0 & (0\%) & 1 & (3\%) & 9 & (28\%) & 10 & (31\%)  \\
Mt\_tRNA &
2 & (9\%) & 6 & (27\%) & 2 & (9\%) & 4 & (18\%) & 3 & (13\%) & 7 & (31\%) & 5 & (22\%) & 18 & (81\%)  \\
miRNA\_pseudogene &
1 & (4\%) & 3 & (14\%) & 0 & (0\%) & 2 & (9\%) & 1 & (4\%) & 1 & (4\%) & 1 & (4\%) & 7 & (33\%)  \\
misc\_RNA\_pseudogene &
2 & (28\%) & 2 & (28\%) & 0 & (0\%) & 0 & (0\%) & 1 & (14\%) & 1 & (14\%) & 2 & (28\%) & 5 & (71\%)  \\
Mt\_rRNA &
1 & (50\%) & 0 & (0\%) & 0 & (0\%) & 0 & (0\%) & 0 & (0\%) & 0 & (0\%) & 1 & (50\%) & 2 & (100\%)  \\
scRNA &
0 & (0\%) & 1 & (100\%) & 0 & (0\%) & 0 & (0\%) & 1 & (100\%) & 0 & (0\%) & 1 & (100\%) & 1 & (100\%)  \\
\hline
totals &2464 & (25\%) & 3418 & (34\%) & 540 & (5\%) & 900 & (9\%) & 1580 & (16\%) & 1661 & (16\%) & 2771 & (28\%) & 7333 & (74\%)  \\
\hline\hline
protein\_coding &
3565 & (7\%) & 4637 & (10\%) & 767 & (1\%) & 1325 & (2\%) & 1190 & (2\%) & 3351 & (7\%) & 4077 & (8\%) & 13751 & (29\%)  \\
\hline
\end{tabular}

\end{table}
\end{landscape}

%
\bibliography{references,gp-bibliography,genechip_papers}

\end{document}